# Validation of the IPSL Venus GCM Thermal Structure with Venus Express Data

Pietro Scarica [1,2,*], Itziar Garate-Lopez [3], Sebastien Lebonnois [4], Giuseppe Piccioni [1], Davide Grassi [1], Alessandra Migliorini [1] and Silvia Tellmann [5]

[1] INAF-IAPS, Istituto di Astrofisica e Planetologia Spaziali, 00133 Rome, Italy; giuseppe.piccioni@inaf.it (G.P.); davide.grassi@inaf.it (D.G.); alessandra.migliorini@inaf.it (A.M.)

[2] Department of Physics, University of Rome Tor Vergata, 00133 Rome, Italy.

[3] Escuela de Ingeniería de Bilbao, Universidad del País Vasco / Euskal Herriko Unibertsitatea, 48013 Bilbao, Spain; itziar.garate@ehu.eus

[4] Laboratoire de Météorologie Dynamique (LMD/IPSL), Sorbonne Université, ENS, PSL Research University, Ecole Polytechnique, Institute Polytechnique de Paris, CNRS, 75252 Paris, France; sebastien.lebonnois@lmd.jussieu.fr

[5] Department of Planetary Research, Rheinisches Institut fr Umweltforschung at the University of Cologne, Aachener Str. 209, 50931 Cologne, Germany; stellman@uni-koeln.de

* Correspondence: pietro.scarica@inaf.it

**Abstract:** General circulation models (GCMs) are valuable instruments to understand the most peculiar features in the atmospheres of planets and the mechanisms behind their dynamics. Venus makes no exception and it has been extensively studied thanks to GCMs. Here we validate the current version of the Institut Pierre Simon Laplace (IPSL) Venus GCM, by means of a comparison between the modelled temperature field and that obtained from data by the Visible and Infrared Thermal Imaging Spectrometer (VIRTIS) and the Venus Express Radio Science Experiment (VeRa) onboard Venus Express. The modelled thermal structure displays an overall good agreement with data, and the cold collar is successfully reproduced at latitudes higher than +/−55°, with an extent and a behavior close to the observed ones. Thermal tides developing in the model appear to be consistent in phase and amplitude with data: diurnal tide dominates at altitudes above $10^2$ Pa pressure level and at high-latitudes, while semidiurnal tide dominates between $10^2$ and $10^4$ Pa, from low to mid-latitudes. The main difference revealed by our analysis is located poleward of 50°, where the model is affected by a second temperature inversion arising at $10^3$ Pa. This second inversion, possibly related to the adopted aerosols distribution, is not observed in data.

## 1. Introduction

Remote-sensing measurements and probe data from space missions [1–4], have been used for decades to investigate the vertical temperature structure of the atmosphere of Venus. Now that the interest towards the planet has been revived, thanks to the Venus Express ESA (European Space Agency) mission [5] (launched in November 2005) and the Akatsuki JAXA (Japan Aerospace Exploration Agency) mission [6] (launched in May 2010), we have the opportunity to perform new investigations based on data acquired with different experiments and techniques.

The understanding of the mechanisms behind the observed features, the dynamics and the structure of Venus' atmosphere, is enhanced by new, modern observations, and numerical simulations conducted by general circulation models (GCMs), that allow us to interpret the physical processes active therein.

GCMs have been extensively used through the years to study Venus' atmospheric circulation. Either relaxing the temperature towards a specified temperature profile or using a radiative transfer module to compute these temperatures, several simulations were already capable of reproducing the observed superrotation [7] in the modelled atmospheres [8–11], even if their results vary from case to case, showing a broad variety of zonal wind fields under similar initial conditions. These results are qualitatively consistent with the well-known superrotation of Venus' atmosphere, which is sixty times faster at the cloud top level than at the surface. GCMs have also been able to demonstrate the role of the thermal tides in the vertical transportation of angular momentum through the atmosphere [12,13], as well as the role of planetary-scale waves and large-scale gravity waves [11,14].

The average temperature field of Venus' atmosphere has been reproduced, from the very extreme physical conditions close to the surface, to the high stability of the mesosphere. However, the atmosphere of the planet has not yet been fully understood and some pending questions still remain unsolved, such as the intriguing polar vortices and the surrounding cold region: observations have unveiled that the polar regions are characterized by variable vortices [15–17], always present, enveloped by a permanent structure known as the cold collar, the temperatures of which are colder towards the morning terminator. Little success was obtained in the modelling of the polar temperature distribution until the work of Ando et al. (2016) [18], based on the AFES model (Atmospheric GCM for the Earth Simulator), where a structured and evolving vortex appeared, along with an inversion of the temperature, located at a level slightly higher than observed [19,20]. The IPSL (Institut Pierre Simon Laplace) simulation presented in Lebonnois et al. (2016) [11] displayed a similar feature around the cloud top level, with temperature variations within the polar regions weaker than observed. However, it was only in the work of Garate-Lopez and Lebonnois (2018) [21] based on the latest version of the IPSL Venus GCM, that an inversion of the temperature was reproduced at the right altitude and latitude. Those authors already noted that temperatures within this cold collar resembling feature, were in good agreement with the observed values of the Visible and Infrared Thermal Imaging Spectrometer (VIRTIS) instrument.

The main goal of the present work is to go into the details of a data-model comparison and, by means of the huge quantity of data available thanks to Venus Express, to validate this version of the IPSL Venus GCM. We focus on comparing the thermal structure reproduced by the GCM developed at the Laboratoire de Météorologie Dynamique of Paris, with the one retrieved by VIRTIS and VeRa (Venus Express Radio Science Experiment), flown on board the ESA Venus Express mission.

General characteristics of the data adopted in this work will be given in Section 2, along with an overview of the state of the art IPSL Venus GCM. The obtained results of this analysis will be presented and discussed in Section 3. At the end, we will summarize and conclude the discussed work in Section 4.

**2. Data and Model**

*2.1. Visible and Infrared Thermal Imaging Spectrometer (VIRTIS) and Venus Express Radio Science Experiment (VeRa)*

Venus Express carried out its investigation of Venus thanks to several instruments [22], spanning from spectrometers to imagers, and from radio occultation to plasma and magnetic field instruments.

The main goal of this work is the comparison of temperature data taken by VIRTIS-M (VIRTIS Spectral Mapper), VIRTIS-H (VIRTIS High-Spectral Resolution) [23] and VeRa [24] with the IPSL Venus GCM output. The thermal structure of Venus atmosphere was also investigated by SPICAV (Spectroscopy for Investigation of Characteristics of the Atmosphere of Venus), but in the 90–140 km range [25], that is marginal for our comparison (performed between 50 and 90 km): for this reason we did not consider these data in the present work. Therefore, among all the instruments onboard, we will just describe the VIRTIS and VeRa experiments, on which data for this study relies. Technical details of the experiments are not the focus of this discussion, although a general overview will be given in this section.

VIRTIS-M and VIRTIS-H, a spectro-imager and a point spectrometer, respectively, are the two subsystems composing VIRTIS. VIRTIS-M is divided in two channels, Visible (VIS) and Infrared (IR), covering a spectral range from 0.3 up to 5.1 μm, with a spectral sampling of about 2 and 10 nm for the two channels, respectively; its projected horizontal resolution per pixel is 16.5 km × 16.5 km at the apocenter distance of 65000 km. VIRTIS-H covers from 2 μm to 5 μm, with a finer spectral sampling, typically equal to 1.5 nm, but variable along the spectral order, and a projected horizontal resolution per pixel of 115 km × 38 km at the apocenter, coarser than VIRTIS-M.

The used temperatures have been retrieved from the $CO_2$ band at 4.3 μm, with a similar methodology applied to both dataset, as reported in the papers of Grassi et al. (2014) [26] (VIRTIS-M) and Migliorini et al. (2012) [27] (VIRTIS-H). Unfortunately, the presence of $CO_2$ non-Local Thermal Equilibrium emission and the atmospheric scattering of the sunlight presently prevent use of this method [26,27] in the illuminated side of the planet, so the data are limited to nighttime only.

The study presented in this work, relies on 636 VIRTIS-M cubes of data acquired until August 2008 [26] and on $3 \times 10^4$ VIRTIS-H spectra collected until November 2013 [27]. The VIRTIS analysis roughly covers from 1 to $10^4$ Pa in both VIRTIS-M and VIRTIS-H cases, which is from the cloud tops, being the lowest level probed by VIRTIS at about 60 km above the surface, up to 90 km, the maximum altitude still effectively probed. This range establishes a limit for our data-model comparison.

VeRa utilizes radio signals in two different bands "X" and "S" (3.6 and 13 cm respectively) to study the Venus atmosphere and ionosphere [24,28]. Measurements have been obtained by directing the High Gain Antenna (HGA) towards the Earth before and after the spacecraft is occulted behind the planetary disc, respectively, as seen from the Earth. The propagation of the radio signal through the ionosphere and the neutral atmosphere of Venus leads to a change in the radio ray path which can be seen as a frequency shift on Earth. This allows retrieval of electron density profiles in the ionosphere and profiles of temperature, pressure and neutral number density in the mesosphere and upper troposphere (~40–90 km) with a high vertical resolution [24]. In the present study, we used 657 VeRa occultations, performed up to November 2013.

Given the orbit geometry of the Venus Express mission, VIRTIS displays in general a better coverage of the whole southern hemisphere with latitudes from mid to high, although VIRTIS-H complements the dataset of VIRTIS-M with more observations at low latitude and in the northern hemisphere.

As happens for VIRTIS, VeRa coverage of the Northern hemisphere is poor too, due to the orbit geometry of Venus Express. However, the radio occultation technique allows to sound deeper below the clouds, probing altitudes down to 40 km [29]. It has already been noted in Limaye et al. (2017) [30], that below 70 km, VIRTIS temperatures are lower than those derived by VeRa, either due to spatial and temporal variations between observations or due to the different techniques used in each study: radio occultations are performed at microwave wavelengths, which are not sensitive to the clouds in the atmosphere of Venus, in contrast to VIRTIS.

The better spatial coverage of VIRTIS and the better accuracy of VeRa measurements, can be used as complementary tools: for these reasons, a joint comparison of VIRTIS and VeRa with the IPSL Venus GCM is interesting.

We limit our analysis to the southern hemisphere, which is more rich in observational data. On the other hand, we would expect Venus to be a symmetric planet (little axis inclination, no seasonal variations, no macroscopic topographic asymmetries between hemispheres, except for the Maxwell Montes) and we can extend, at least qualitatively, any consideration done for the southern hemisphere to its northern counterpart. Also, because the model itself would eventually advise us about any possible asymmetry of the planet: the northern and the southern hemisphere are almost identical in the IPSL Venus GCM.

## 2.2. The Institut Pierre Simon Laplace (IPSL) Venus General Circulation Model (GCM)

The dynamical core of the IPSL Venus GCM is based on the Earth model developed at the Laboratoire de Météorologie Dynamique of Paris [31], which is a latitude-longitude grid finite-difference dynamical core. The model has the capability to zoom over a given region.

The simulation in this work utilized a grid of 96 longitudes × 96 latitudes, which is a horizontal resolution of 3.75° × 1.975° with 50 hybrid vertical levels (from 0 to 95 km altitude). This set up and simulation was originally presented in Garate-Lopez and Lebonnois (2018) [21].

The IPSL Venus GCM uses the boundary layer scheme of Mellor and Yamada (1982) [32], and a temperature-dependent specific heat $C_p(T)$, in order to get realistic adiabatic lapse rates in the entire atmosphere. It is also provided with a realistic topography taken from the Magellan mission.

Instead of characterizing the circulation of Venus atmosphere through simplified heating and cooling, as well as a rough representation of clouds, the IPSL model counts on a full radiative transfer module, that computes the temperature structure self-consistently. This radiative transfer module takes into account the cloud model of Haus et al. (2014, 2015) [33,34], that is based on Venus Express observations. This cloud model gives the vertical distribution for the three cloud particle modes present in Venus clouds (radius near 0.1 μm for mode 1, 1 (1.5) μm for mode 2 (2') and 3 μm for mode 3). It includes a latitudinal modulation of these distributions [33]: the strong decrease of the cloud top altitude at 50° latitude, dropping from 70 km to 61 km over both poles, and a latitudinally dependent scaling of the abundance of the different modes compared to the equatorial vertical distributions.

For the IPSL Venus GCM radiative transfer module, solar heating rates are computed from look-up tables that give the heating rates as a function of altitude, solar zenith angle, and latitude. This cloud model is also implemented in the infrared net exchange matrices used by the IPSL model. As the simulated temperature structure in the deep atmosphere was colder (10 K) than that observed, we increased the solar heating rates in the 30–48 km altitude region by multiplying the values provided by Haus et al. (2015) [34] by a factor of 3. The composition of the lower haze particles located in this altitude range is not well established yet and, therefore, the optical properties and the absorption of the solar flux in this region are subject to uncertainty. This allows some tuning in the radiative transfer module to bring the modelled temperature profile in agreement with the observed values.

In order to implement the latitudinal variation of the cloud structure in the infrared cooling rates, the net-exchange rate matrices [35] are computed for five latitudinal bins and then interpolated between the central latitudes of each bin. These matrices use correlated-k coefficients of opacity sources, gas and clouds, and take into account the $CO_2$ and $H_2O$ collision-induced absorption [36].

Some additional continuum (i.e., extinction coefficient set to a minimum value) to close the infrared windows located in the 3–7 μm range was also needed below the clouds (16 to 48 km) in order to have a best fit of the VIRA [37] and VeGa-2 [38,39] temperature profiles.

More details about the simulation used for this study and the main characteristics of the radiative transfer scheme, can be found in Garate-Lopez and Lebonnois (2018) [21], while an accurate description of the IPSL Venus GCM is given in Lebonnois et al. (2010) [13].

## 3. Results and Discussion

In order to validate the global thermal characterization of the Venus atmosphere obtained with the IPSL Venus GCM, we first analyze the zonally and temporally averaged temperature field and then the main variations with respect to that mean, i.e., the thermal tides, by comparing them with the structures obtained from observations made by Venus Express. The other types of planetary-scale waves present in the atmosphere are not analyzed in the present work.

*3.1. The Thermal Field*

The simulated temperature field is easy to compare with Venus Express observations and it gives an immediate idea of the model reliability. We performed a zonal and temporal average of the temperature field. The average has been made over 360° of longitude and the entire night (from 18:00 to 6:00 in local time), for both model (using a simulation through 2 Venus days) and data (using the entire available dataset). The comparison has been produced for the two hemispheres; here is just reported the discussion for the—better observationally covered—southern part of the globe.

The altitude-latitude distribution of the modelled temperature field (Figure 1), shows a flat trend in the low-to-mid latitude range, around $10^3$ Pa. This pressure level represents the border between two different dynamical regions in the modelled temperature distribution. Above this level, temperature increases from the equator to the south pole; on the contrary, below this level, temperature increases towards the equator.

In general, equatorward of −60°, the temperature field seems to be homogeneously stratified, no complex features are present and no abrupt variation of the temperature. Moreover, in this region, the temperature trend is monotonically decreasing from the bottom to the top of the atmosphere. On the contrary, at higher latitudes a more complex behavior is modelled. There are two cold elongations extending downward and poleward from −60°; one at pressures higher than $4 \times 10^3$ Pa and another at pressures lower than $2 \times 10^3$ Pa. The former resembles the observed cold collar not yet being completely isolated [21], and shows temperatures 15–20 K colder than the surroundings. The latter is coupled with a warmer region at high latitudes which extends almost 10 Pa into the vertical in the simulated temperature field. The absolute temperature of the upper cold and warm regions, as well as the thermal contrast they maintain with their surroundings, is lower than that of the regions at ~$10^4$ Pa (64 km).

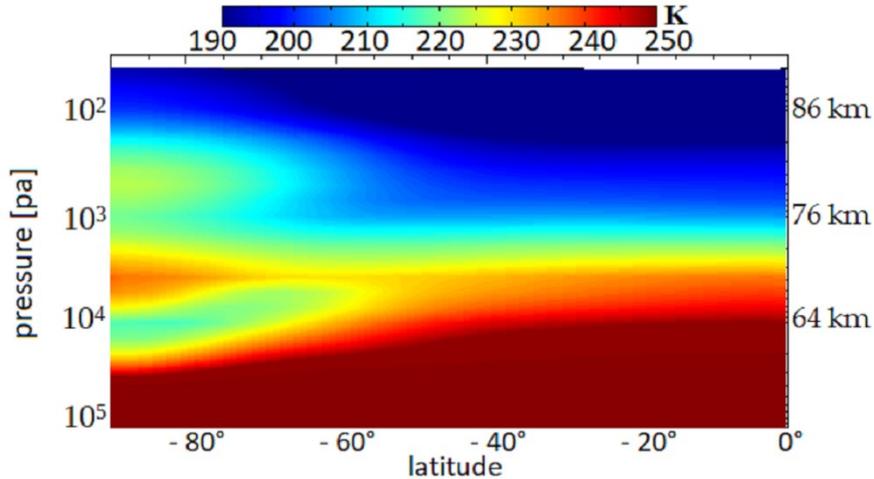

**Figure 1.** Altitude–latitude distribution of the zonal and temporal average temperature field (K) for the Institut Pierre Simon Laplace (IPSL) Venus general circulation model (GCM) simulation. The average has been obtained over 360° of longitude and the entire night.

The zonal and temporal average temperature maps, based on observations by VIRTIS-M (Figure 2a), VIRTIS-H (Figure 2b) and VeRa (Figure 2c), show a qualitative agreement with the IPSL Venus GCM, with the exception of very high latitudes (poleward of −80°) where, especially in VeRa data, it is possible to recognize a fully formed warm pole, more vertically extended than the one reproduced by the model. The general contrast between the warm and cold areas close to the pole in the data-based maps and in the simulated one, is very similar.

At $10^3$ Pa, the data distributions display a latitudinal upward trend from the equator to the pole, particularly clear for VIRTIS-M and VeRa, where the latitudinal profile is monotonically changing. No abrupt variations—like that present in the model at −70°—are observed in the high atmosphere of VIRTIS and VeRa data. Beside of these differences, we can confirm the overall similarity of model and data altitude-latitude distributions: below and above the $10^3$ Pa pressure level we can recognize, in all the observations-based maps, the two dynamical regions that we see in the GCM. Moreover, in the deep atmosphere, at altitude roughly below $3 \times 10^4$ Pa, the model is very much in agreement with VeRa data. Even more striking, the major feature reproduced by the GCM and associated to the cold collar is observed in VIRTIS-H, VIRTIS-M and VeRa at very similar pressure levels (roughly below 3–4 × $10^3$ Pa) and latitudes (poleward of −60°). The modelled temperatures within this region are close to those that are observed.

It is clear from this first comparison that the most important discrepancies between the IPSL Venus GCM simulation and Venus Express observations regarding the mean temperature field are the cold elongation around $10^3$ Pa and the vertical extension of the warm region associated with it.

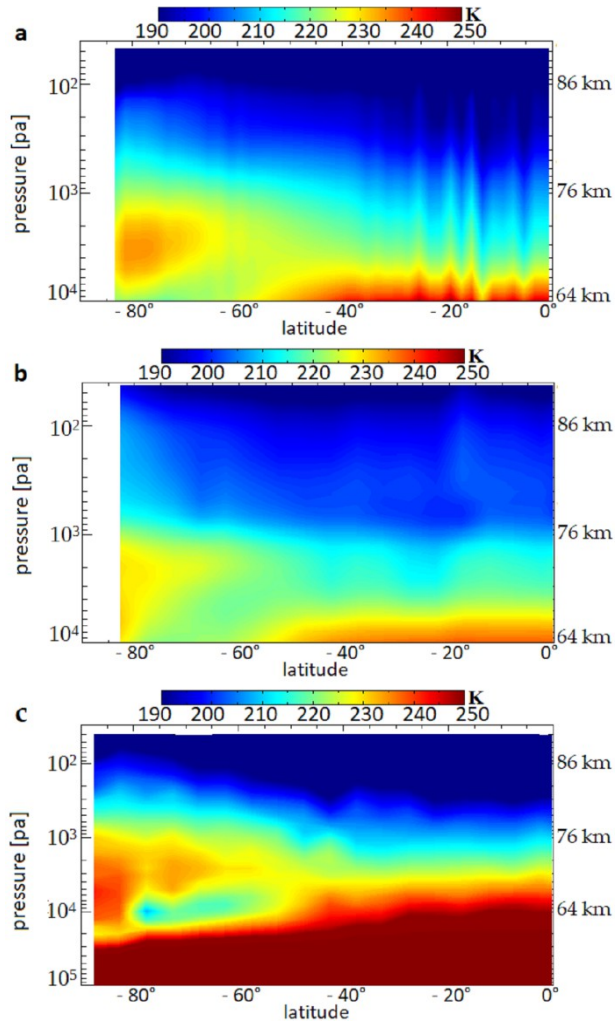

**Figure 2.** Altitude-latitude distribution of the zonal and temporal average temperature field (K). (**a**) VIRTIS-M (VIRTIS Spectral Mapper); (**b**) VIRTIS-H (VIRTIS High-Spectral Resolution); (**c**) Venus Express Radio Science Experiment (VeRa). The average has been obtained over 360° of longitude and the entire night.

In order to test more thoroughly the vertical temperature gradient simulated by the GCM at high latitudes, we built a latitude-by-latitude comparison of GCM and data average temperatures and compared the high-latitude profiles with low and mid latitude samples. The averages have been obtained over the entire nighttime and 360° of longitude, for both model and data. Here, we show a set of ad-hoc reference latitudes (−75°, −45° and −15°), which are characteristic of high, mid and low latitude sample (Figure 3).

At −75° (Figure 3a), the average temperature in the simulation peaks at 235 K on a pressure level of 3–4 × $10^3$ Pa (around 69 km). This is consistent with VIRTIS-M and VeRa profiles. The peak in VIRTIS-H data is located at a slightly lower temperature (230 K) and at a lower pressure level (2 × $10^3$ Pa) than GCM output, VIRTIS-M and VeRa data.

Deeper in the atmosphere (between 4 × $10^3$ Pa and 2 × $10^4$ Pa), model and data share the clearly recognizable cold collar's temperature inversion: after peaking, the temperature decreases at higher pressure levels. The contrast between the peak and the deepest region of the cold collar is about 15 K in the model, in agreement with VIRTIS-M and VeRa data. The transition between the warm layer above the cold collar and the cold collar itself, takes place at a shorter vertical distance in the model: thus, the peak of the profile is sharper in the GCM than in data.

At this latitude, the GCM displays a second temperature inversion located between 5 × $10^2$ Pa and 2 × $10^3$ Pa, coinciding with the upper cold elongation discussed in the altitude-latitude distribution maps. This feature does not appear in either VIRTIS-M or VeRa data profiles, that are instead characterized by a monotonic increase of the temperature from the top of the observed atmosphere to the peak at 4 × $10^3$ Pa. At pressure levels between $10^2$ and 4 × $10^2$ Pa, VIRTIS-H data display a more flat trend, with respect to VIRTIS-M and VeRa. However, this feature never becomes a complete inversion of the temperature and the double peak structure that appears in the model is far more pronounced than in VIRTIS-H data.

At mid latitude (Figure 3b), model and data average temperature trends show an overall good agreement. As expected, the cold collar disappears in both model and data, and no second upper inversion is produced by the GCM. Actually, this upper inversion present in the model starts as an isothermal feature at around −55° and then evolves into a clear inversion as latitude increases towards the south pole, but shows no sign at all at lower latitudes. At −45° (Figure 3b) both data and GCM profiles display a rising temperature from the top to the bottom of our accessible pressure range. In the pressure range between 3 × $10^3$ Pa and $10^4$ Pa, GCM temperatures are 5–10 K warmer than data; deeper in the atmosphere, the only available observations are VeRa data. At low pressures (below 2 x $10^2$ Pa), the discrepancy of model and data profile is about 10 K, slightly more than the random errors due to the retrieval method, that are about 4–5 K in this pressure range [26]. Moreover, we have to consider that the strong variability of the Venus atmosphere, also in terms of temperature profiles, may be important for short-term evolution and utilizing poor statistical coverage data in the average. We note that VeRa data make an exception and they are very much in agreement with the model in the entire range of analysis, apart from the pressure levels between 3 × $10^3$ Pa and 3 × $10^4$ Pa.

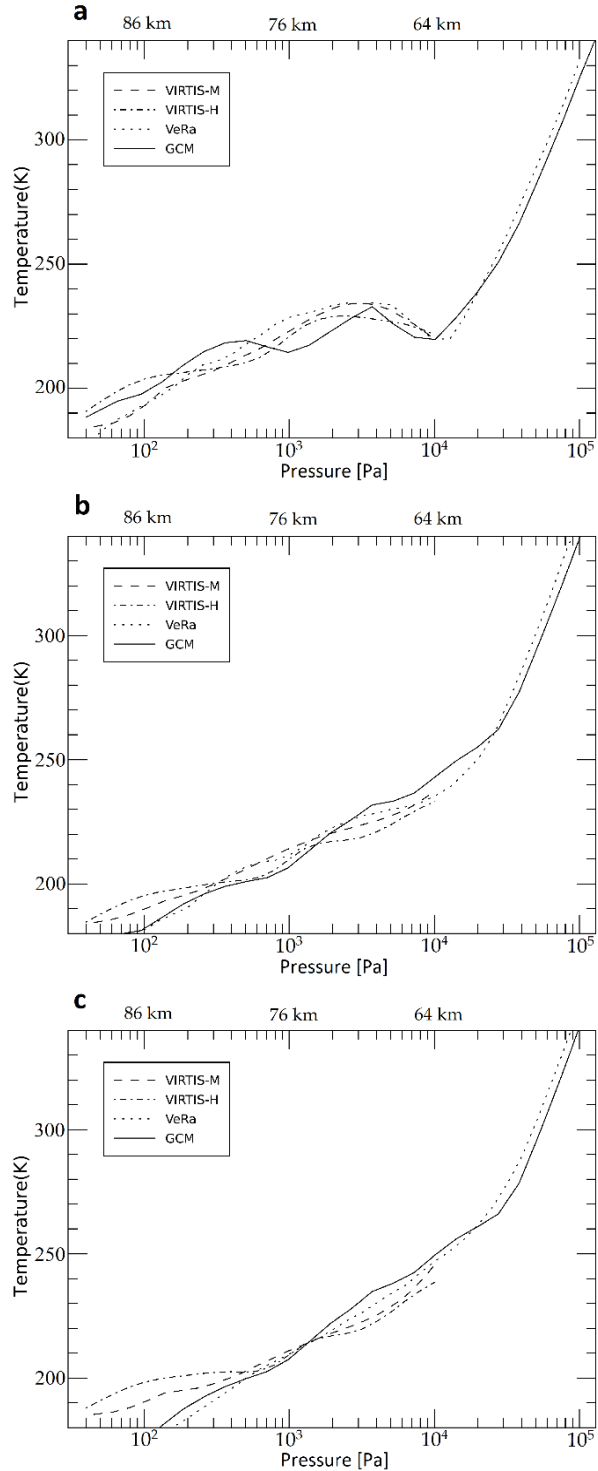

**Figure 3.** Average temperature profiles (K) for the IPSL Venus GCM (solid line), VIRTIS-M data (dashed line), VIRTIS-H data (dotted-dashed line) and VeRa data (dotted line), at: (**a**) −75°; (**b**) −45°; (**c**) −15°. The average has been obtained over 360° of longitude and the entire night.

Moving towards the equator, at −15° (Figure 3c), a very good consistency of model and data profiles is seen in the mid-pressure range (5 × 10$^2$–2 x 10$^3$ Pa), where random retrieval errors in VIRTIS-M and VIRTIS-H are smaller (about 1 K for both VIRTIS-M and VIRTIS-H, [26,27]). On the sides of the lowest and highest pressure levels of the atmosphere probed by VIRTIS, a broad difference is visible between the GCM average temperature profile and VIRTIS data (both -M and -H), up to 15–20 K around $10^2$ Pa. This large discrepancy may be partially due to the combined effect of the low spatial coverage of VIRTIS at low latitudes and the intrinsic error of the retrieval method at high and low pressures. However, the GCM profile remains consistent with the VeRa one through the entire pressure range. Moreover, while the upper inversion is not present in the model, in VIRTIS-M and VeRa data, we still notice a flattened trend in the VIRTIS-H profile, at pressures below $5 \times 10^2$ Pa. This feature is shown for the entire southern hemisphere of VIRTIS-H data, and thus is not related to the mid-to-high latitude upper inversion that we see in the model.

We now inspect the horizontal structure of both thermal inversions using polar projections. Figure 4 shows the nighttime temperature field of the southern hemisphere at $7 \times 10^3$ Pa, where the cold collar resides, as observed by VIRTIS-M data (Figure 4a) and simulated by the model (Figure 4b). The radial coordinate of the grid corresponds to latitude and the angular coordinate to local time, going from the evening to the morning terminator clockwise. The model temperature field in Figure 4b is an average over 2 Venus days, while the VIRTIS-M temperature field in Figure 4a is obtained with all available data.

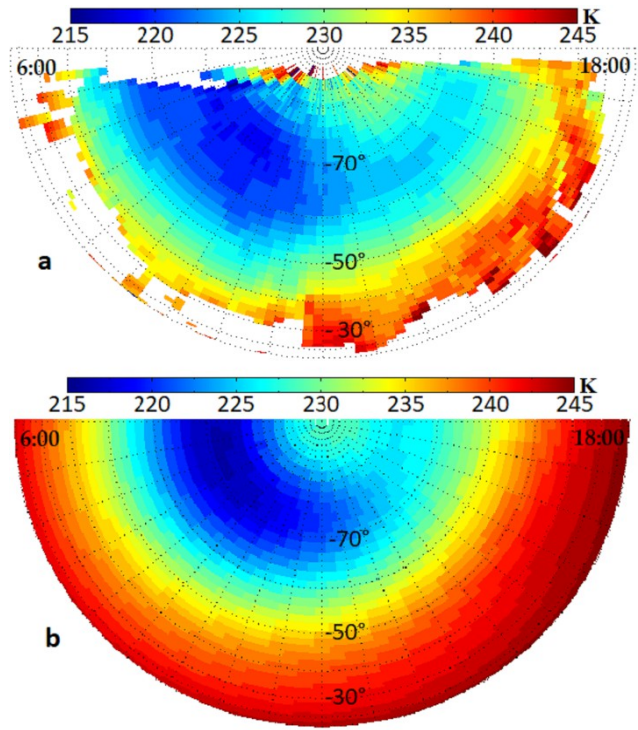

**Figure 4.** Nighttime polar temperature field (K) at $7 \times 10^3$ Pa for (**a**) VIRTIS-M and (**b**) the IPSL Venus GCM simulation.

Although there is a lack of observations close to the equator and to the morning and evening terminators, it is clear that the GCM is consistent with data: the first half of the night is in general warmer than the second one, in both the simulation and VIRTIS-M data, and local minima and maxima that appear in the model are consistent with the data. The cold collar in the IPSL Venus GCM is a completely formed structure, well distinguished from the surrounding environment. The feature developing in the model displays a strong resemblance with the data: the altitude where the feature is peaking (around 64 km, roughly $10^4$ Pa) and the temperature values (about 220 K) obtained in the simulation are consistent with observations, as well as the phase in local time. Both GCM and VIRTIS-M cold collars are about 20 K colder than low latitudes and are more pronounced after midnight, in particular towards the morning terminator. The GCM cold collar shows its maximum at latitudes slightly higher than −70°, the observed one a few degrees equatorward, about −65°.

Figure 5 shows the comparison at $2 \times 10^3$ Pa pressure level (corresponding to the upper temperature inversion), where the observed nighttime temperature fields (Figure 5a) and the simulated (Figure 5b) display clear differences. We already know, from the altitude–latitude distribution that, at this altitude, VIRTIS-M shows a monotonic increasing regime: the temperatures rise from the equator to the pole, with retrieved values spanning in a range of 10–15 K. While the temperature range in the model is in general similar to that observed, the horizontal distribution is completely different. In the GCM output, temperatures become colder from the equator to mid-latitude, and then starts to become warmer and warmer poleward. Temperature values around −80° are similar in the GCM and VIRTIS-M fields, therefore, the transition between mid and high latitudes in the second half of the night is sharper in the simulation than in the observations. This modelled upper inversion shares some characteristics with the cold collar: it just appears in the second half of the night and it peaks around −60°. In the GCM map a warm area is visible in the first part of the night, close to the evening terminator; in the VIRTIS map, this region also appears to be warmer, but it is not clear enough to draw conclusions.

The thermal structure of Venus atmosphere has been studied previously [20,40,33] and all these works showed the mentioned general trend, the two different dynamical regions above and below $10^3$ Pa, and a cold region with its thermal inversion, but none of these temperature maps obtained from observations showed the upper inversion. It is worth mentioning that Ando et al. (2016, 2017) [18,41], obtained with the AFES Venus GCM a thermal structure similar to our upper inversion. They identified it as the cold collar. However, comparing their results with Figure 1 in this work, we conclude that their cold elongation, and therefore thermal inversion, is more similar to our upper inversion than to the inversion corresponding to the cold collar, at least regarding the pressure vs latitude structure.

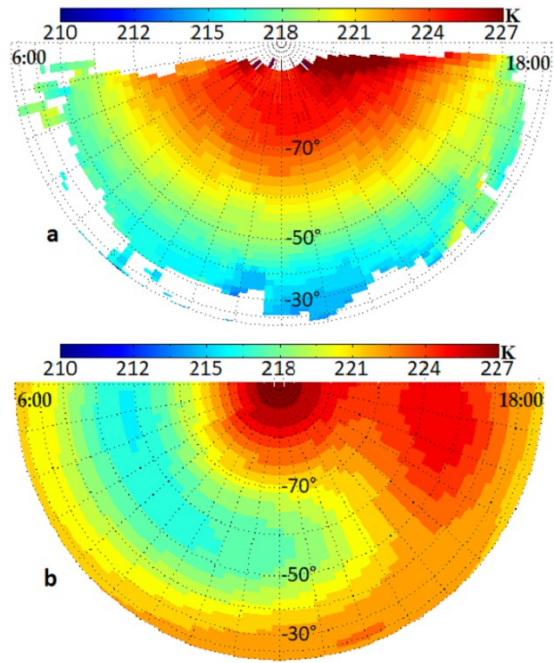

**Figure 5.** Nighttime polar temperature field (K) at $2 \times 10^3$ Pa (roughly corresponding to 72 km) for (**a**) VIRTIS-M and (**b**) the IPSL Venus GCM simulation.

*3.2. The Thermal Tides*

In order to evaluate the temporal variation of the temperature, altitude-local time distributions of temperature anomalies (defined as the temperature differences to the local time averaged over the entire night) are shown in Figures 6 and 7 for the GCM simulation and for VIRTIS observations. These plots allow the visualization of the thermal tides induced in the atmosphere by the diurnal cycle of the solar heating rates.

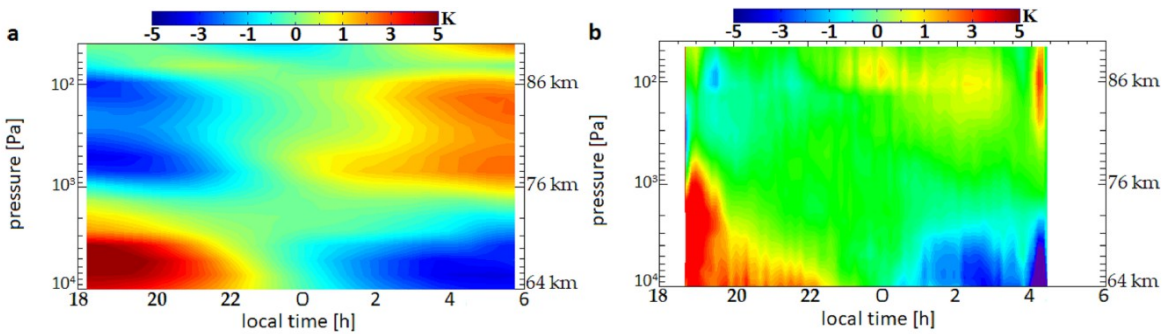

**Figure 6.** Altitude-local time distribution of the temperature residuals anomalies (K) at –75°. (**a**) GCM simulation; (**b**) VIRTIS-M observations.

For the high-latitude and mid-latitude samples a robust comparison can be made, which is not the case for the low latitude one, where VIRTIS data coverage is not enough to produce meaningful maps. The same is true for VIRTIS-H at high latitude. The right edge of the maps in Figures 6 and 7 corresponds to the dawn terminator, while the left ones to the evening terminator. Data close to the dawn terminator have been removed in our analysis, due to the presence of scattered sunlight coming from daytime.

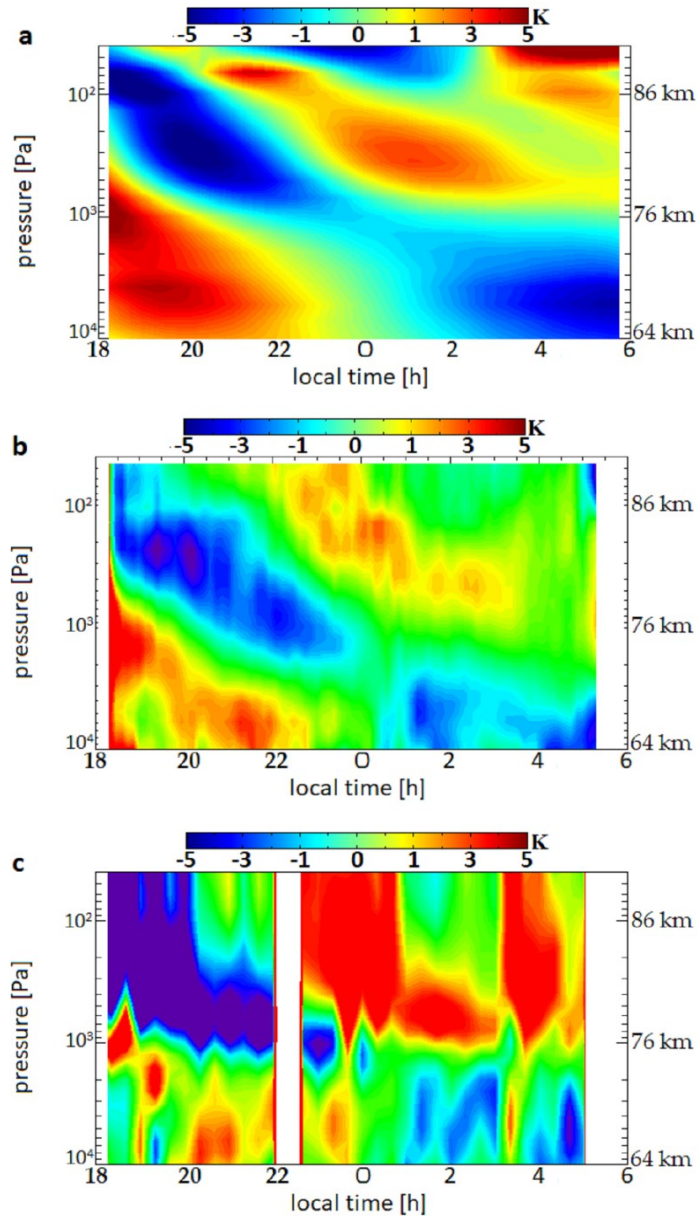

**Figure 7.** Altitude-local time distribution of the temperature residuals anomalies (K) at −45°. (**a**) GCM simulations; (**b**) VIRTIS-M observations; (**c**): VIRTIS-H observations.

Figure 6 shows the temperature anomalies' maps at −75° latitude. Model and data share two clearly recognizable regimes: below $10^3$ Pa, a warmer first half of the night (until 22:00) and a colder second half (after 0:00), and an inverted tendency above $10^3$ Pa. This warm–cold dichotomy, that seems to be mainly related to the diurnal tide, is thus an evidence of the GCM capability to reproduce the observed thermal tides. Although both the simulation and the observations show a quite similar general behavior, cold and warm regions in the high atmosphere are less pronounced in VIRTIS-M data. Moreover, while in the model these maxima and minima extend from $10^2$ Pa to $10^3$ Pa, in VIRTIS-M data they just appear in a narrower region around $10^2$ Pa.

At −45° (Figure 7) the maximum-minimum couple above $10^3$ Pa splits up into a component with half of the period, which phase propagates into the higher layers as we move towards the evening terminator. At low pressure levels, the semidurnal tide is clearly dominating in the model, while the diurnal tide is still recognizable below $10^3$ Pa, but it slowly fades moving towards the equator (not shown). The appearance of a dominating semidiurnal component at mid latitudes is confirmed by both VIRTIS-M and VIRTIS-H data, with a striking resemblance with the GCM. Thus, the modelled thermal tides seems to be in good agreement with the observed ones, at all latitudes. The amplitude is consistent with observations (about 5 K at cloud top), as well as the phase in local time.

Note that Figures 6 and 7 show the total temperature deviations (thus showing the dominant tides at each latitude), and not the thermal tides' wave number 1 and 2 components (discussed below). However, it seems that the majority of the temperature anomalies comes from diurnal and semidiurnal tides, and that higher harmonics are much less significant when contributing to the total temperature deviations.

Latitude-pressure maps of the fast Fourier transform (FFT) diurnal and semidiurnal components of the modelled temperature are presented in Figure 8a and Figure 8b, respectively; they illustrate the regions where these tides are dominating and they give information about the amplitude of the two main components.

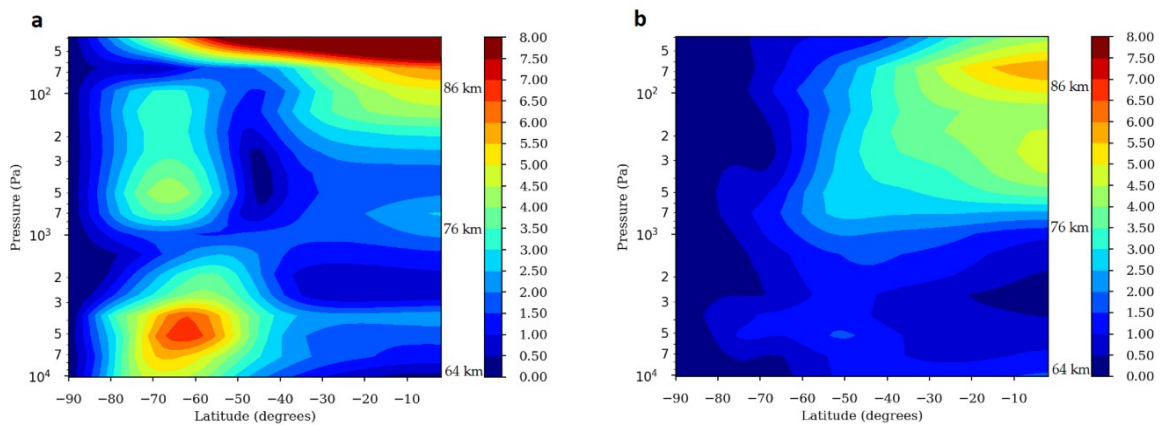

**Figure 8.** Fast Fourier transform (FFT) components of the temperature in the GCM simulation: (**a**) diurnal component; (**b**) semidiurnal component.

Ando et al. (2016) [18] suggested that the thermal tides are crucial for the structure of Venus upper polar atmosphere and above cloud levels. In order to study this possibility, we analyzed the horizontal structure of the temperature anomaly field and its three main harmonics (Figure 9), for a pressure level around $5 \times 10^2$ Pa, roughly corresponding to the upper inversion level. The quarterdiurnal component have not been reported in this study; its amplitude is much smaller than all the other main harmonics, along the entire hemisphere. The horizontal structure of the overall temperature anomaly field (Figure 9a) shows that thermal tides obtained in the IPSL Venus GCM have a large impact from the equator until −80° latitude. The diurnal component (Figure 9b) is the one responsible for the thermal tides signature at high latitudes; its amplitude peaks between −55° and −80° and has a value of 4 K. The semidiurnal component (Figure 9c) dominates at latitudes equatorward of −55° and peaks at the equator with an amplitude of 5 K. The terdiurnal component (Figure 9d) reaches its maximum amplitude of 3 K at mid-latitudes (between −30° and −60°).

This behavior is consistent with our previous simulations [11] but shows some differences with Ando et al. (2018) [42]. They produce a very similar horizontal structure of temperature anomalies (total, diurnal, and semidiurnal deviations) at 80 km altitude. Nevertheless, their maximum values for the total temperature deviation are somewhat higher while the amplitudes of the diurnal and semidiurnal tides for both studies are comparable, which suggests that higher wave number components could be significant in their model. In fact, Takagi et al. (2018) [43] studied the amplitudes of the terdiurnal and quarterdiurnal tides in the AFES-Venus GCM simulations, and showed that in the vertical winds, peak amplitudes of these higher harmonics are roughly 50% and 25%, respectively, compared to semi-diurnal tide. In our FFT analysis of the simulated temperature field, the terdiurnal component appears to be also significant, in particular near mid-latitudes, as shown in Figure 9, while the quarterdiurnal component seems negligible.

Figure 9 indicates that the most significant component in the upper inversion area ($5 \times 10^2$ Pa, 60°–90°) is the diurnal tide, whose maxima and minima locate around −65° latitude (Figure 9b). However, Ando et al. (2018) [42] produce these maxima and minima at 45°, out of the upper inversion area. Consequently, due to the different thermal tide contributions, we find it unlikely that the upper thermal inversion seen in both the IPSL Venus GCM and the AFES-Venus GCM is due to the thermal tides in this area.

Because it is a thermal structure, the origin of the upper inversion is certainly related to the heating or cooling rates used in each GCM. According to Ando et al. (2018) [42], in AFES-Venus GCM solar heating rates are based on observations by Tomasko et al. (1980) [44], and moving upwards the top altitude of the solar heating from 80 km to 90 km did not affect the thermal structure much. In the IPSL model (as detailed in Section 2) the solar heating rates are based on the radiative transfer model by Haus et al. (2015) [34]. In Garate-Lopez and Lebonnois (2018) [21], it was shown that compared to Lebonnois et al. (2016) [11] who were using solar heating rates based on Crisp (1986) [45], changing only the solar heating rates did not affect the polar thermal structure near the cloud-top. Therefore, it seems that this upper inversion is barely sensitive to the solar heating computation scheme. This is consistent with the fact that at latitude poleward of 60°, the thermal balance is mostly obtained between dynamical heating and infrared cooling. For the infrared cooling rates, using the latitudinal variation in the cloud structure did affect significantly the polar thermal structure in the work of Garate-Lopez and Lebonnois (2018) [21], inducing the presence of the thermal inversion similar to the cold collar. However, it reduced the upper inversion, but it did not remove it, as discussed in this work.

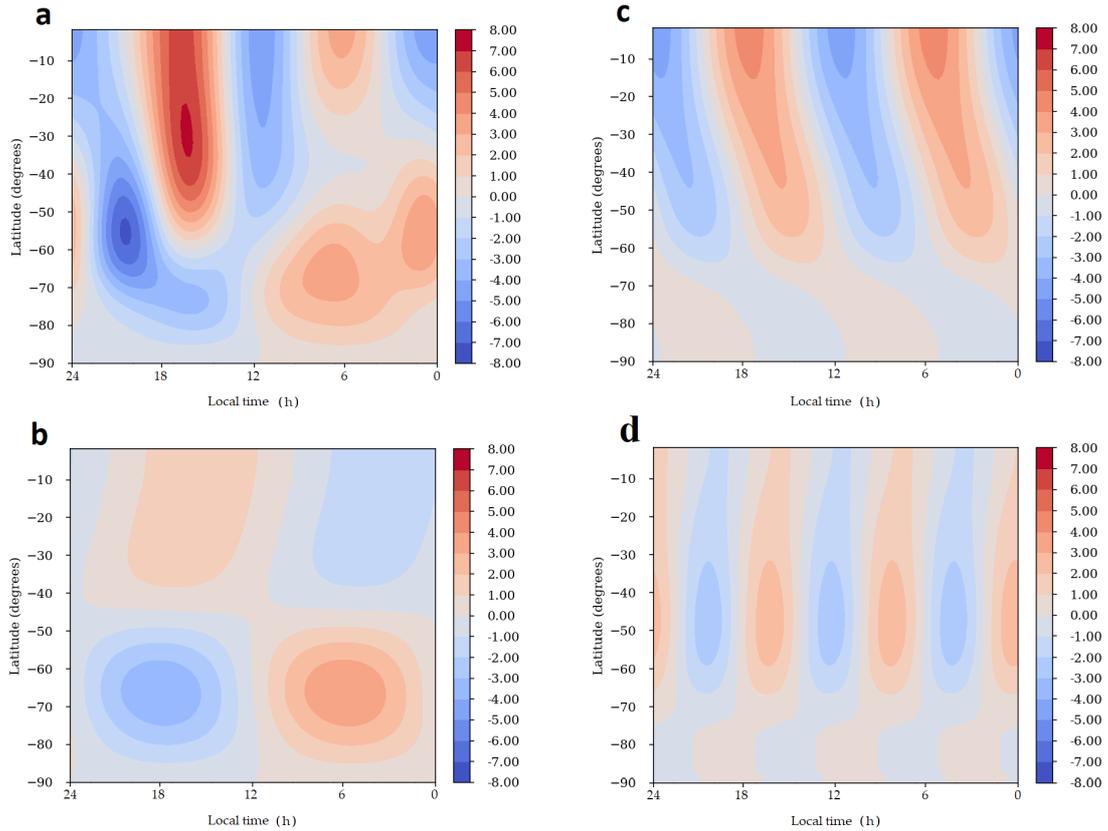

**Figure 9.** Horizontal structure at $5 \times 10^2$ Pa (80 km): (**a**) temperature anomaly field (K); (**b**) first harmonic of the temperature anomaly field (K); (**c**) second harmonic of the temperature anomaly field (K); (**d**) third harmonic of the temperature anomaly field (K).

Both GCMs have slightly different vertical resolution for the calculation grid, 1 km for the AFES-Venus [42] and 1.5 km for the IPSL Venus GCM around the upper inversion's level. Although we do not believe that vertical resolution is responsible for this structure, it could help to unmask its nature. Therefore, a new simulation with better vertical resolution, at least in the cloud region and above, would be of interest for future work.

From our point of view, it is most likely that the origin of this thermal structure is related to the particle distribution used in the model in the region of this upper inversion. The latitudinal variations suggested by the work of Haus et al. (2015) [34] may not be enough to model the balance between radiative cooling and dynamical heating correctly. Unfortunately, observations of the opacity, composition and particle size in the clouds remain incomplete, especially in the polar atmosphere. Testing the sensitivity of the thermal structure to the distributions of mode 1 and mode 2 particles (both present in the whole cloud deck) may be done with the IPSL Venus GCM, though this effort is beyond the scope of this paper.

## 4. Summary

Venus' temperature fields—obtained by VIRTIS and VeRa experiments onboard Venus Express—have been compared to the latest version of the IPSL Venus GCM, in order to validate the model and to better understand how the Venus' atmosphere works.

The zonal and temporal average of the simulated temperature field displays a similar temperature range and a general agreement with observations: both data and model have two different dynamical regions, with temperatures increasing from the equator to the south pole above the $10^3$ Pa pressure level, and decreasing towards the pole below the same level. At pressures higher than $3–4 \times 10^3$ Pa and poleward of −60°, the model reproduces a well formed cold feature that resembles the observed cold collar.

The unsolved question is the presence of an upper cold elongation located above $2 \times 10^3$ Pa, coupled with a warmer region at high latitudes. This cold feature, already present in the former version of the IPSL Venus GCM, as well as in the AFES model, does not change with respect to the old simulations, in both IPSL and AFES, despite of modifications that have been made in the rates calculation schemes between previous and current versions of the two models [11,13,18,21,41,42].

The analysis of long term temperature variations, reveals a high latitude warm-cold dichotomy of $1/V_d$ (1 cycle per Venus day) frequency and a low-mid latitude $2/V_d$ component that arises in the low pressures (below $10^3$ Pa). The comparison of the model with VIRTIS data shows a very good agreement in both amplitude and phase. The larger contributions in the IPSL Venus GCM temperature anomalies are due to the first and the second harmonic of thermal tides, with the diurnal tide that is the most important component at high latitudes (maxima centered at $5 \times 10^2$ Pa and $5 \times 10^3$ Pa), and the semidiurnal tide that dominates at low and mid latitudes (between $10^2$ and $10^3$ Pa).

According to our analysis, the horizontal structure of the temperature anomaly field in the IPSL Venus GCM shows some discrepancies with that obtained by Ando et al. (2018) [42]. Diurnal tide maxima and minima are located around −65° in our model and around 45° in Ando et al. (2018) [42]. Moreover, the comparable amplitudes of the diurnal and semidiurnal tides of the two studies, despite of the discrepancies in the amplitudes of the total temperature deviation, implies a larger contribution of higher wave number components in their model. Due to these reasons, we find no hint of a thermal tides involvement in the formation of the upper cold elongation: it is produced in both the IPSL Venus GCM and the AFES-Venus GCM, despite the discrepancies in the simulated thermal tides.

Our knowledge of the particle distribution in the Venus polar atmosphere, on which the aerosol density profiles that we used in our model relies [34], is still incomplete. We suggest that this upper cold elongation is due to the distribution of the particles, meaning that the aerosol density profile used in our model probably does not accurately represent the actual atmosphere of Venus in the polar regions above the clouds. We believe this lack of accuracy, significantly affecting the cooling rates, could be the cause of the cold elongation, more than thermal tides' contribution or heating rates calculation.

Author Contributions: conceptualization, P.S., I.G-L, S.L. and G.P.; methodology, I.G-L, S.L., G.P., D.G., A.M. and S.T.; software, P.S.; validation, P.S., I.G-L and S.L.; formal analysis, P.S.; investigation, P.S.; resources, I.G-L, S.L., G.P., D.G., A.M. and S.T.; data curation, P.S.; writing—original draft preparation, P.S.; writing—review and editing, I.G-L, S.L., G.P., A.M. and S.T.; visualization, P.S.; supervision, I.G-L, S.L. and G.P.; project administration, G.P.; funding acquisition, G.P.

**Acknowledgments:** The GCM simulations were done thanks to the High-Performance Computing (HPC) resources of Centre Informatique National de l'Enseignement Supérieur (CINES) under the allocations n° A0020101167 and A0040110391 made by Grand Équipement National de Calcul Intensif (GENCI). SL acknowledges financial support from the Centre National d'Etudes Spatiales (CNES), and from Programme National de Planétologie (PNP) of CNRS/INSU, co-funded by CNES.

This research has been supported by ASI (Agenzia Spaziale Italiana) and INAF (Istituto Nazionale di Astrofisica).

**Conflicts of Interest:** The authors declare no conflict of interest.